\documentclass[preprint, 12pt]{aastex}

\usepackage{graphicx}
\usepackage[space]{grffile}
\usepackage{latexsym}
\usepackage{textcomp}
\usepackage{longtable}
\usepackage{multirow,booktabs}
\usepackage{amsfonts,amsmath,amssymb}
\usepackage{natbib}
\usepackage{url}
\usepackage{hyperref}
\hypersetup{colorlinks=false,pdfborder={0 0 0}}
\usepackage[utf8]{inputenc}
\usepackage[english]{babel}

\newcommand{\ic}{IC 10 X--1}
\newcommand{\spin}{a_*}
\newcommand{\mdot}{{\dot{M}}}

\newcommand{\msun}{{\rm M}_\odot}
\newcommand{\Msun}{{\rm M}_\odot}

\newcommand{\chandra}{{\it Chandra}}
\newcommand{\nustar}{{\it NuSTAR}}

\newcommand{\xmm}{{\it XMM-Newton}}

\slugcomment{To appear in The Astrophysical Journal (ApJ)}

\shorttitle{On the Spin of the Black Hole in IC 10 X-1}
\shortauthors{Steiner et al.}

\begin{document}


\title{On the Spin of the Black Hole in IC 10 X-1}


 \author{
  James F.\ Steiner\altaffilmark{1,2}\altaffilmark{\dag},
  Dominic J.\ Walton\altaffilmark{3},
  Javier A.\ Garc{\'{\i}}a\altaffilmark{2},
  Jeffrey E.\ McClintock\altaffilmark{2},\\
  Silas G.\ T.\ Laycock\altaffilmark{4}, 
  Matthew J.\ Middleton\altaffilmark{5},
  Robin Barnard\altaffilmark{2},
  Kristin K.\ Madsen\altaffilmark{6}
}

\altaffiltext{1}{MIT Kavli Institute for Astrophysics and Space  Research, MIT, 70 Vassar Street, Cambridge, MA 02139, USA.}
\altaffiltext{2}{Harvard-Smithsonian Center for Astrophysics, 60  Garden Street, Cambridge, MA 02138, USA.}
\altaffiltext{3}{NASA Jet Propulsion Laboratory, 4800 Oak Grove Dr., Pasadena, CA 91109, USA.}
\altaffiltext{4}{Department of Physics, University of Massachusetts Lowell, MA 01854, USA.}
\altaffiltext{5}{Institute of Astronomy, Madingly Road, Cambridge, CB3 OHA, UK.}
\altaffiltext{6}{Cahill Center for Astronomy and Astrophysics, California Institute of Technology, Pasadena, CA 91125, USA.}
\altaffiltext{\dag}{Einstein Fellow.}

\begin{abstract}

The compact X-ray source in the eclipsing X-ray binary IC 10 X-1 has reigned for years as ostensibly the most massive stellar-mass black hole, with a mass estimated to be about twice that of its closest rival. However, striking results presented recently by Laycock et al. reveal that the mass estimate, based on emission-line velocities, is unreliable and that the mass of the X-ray source is essentially unconstrained. Using {\it Chandra} and {\it NuSTAR} data, we rule against a neutron-star model and conclude that IC 10 X-1 contains a black hole.  The eclipse duration of IC 10 X--1 is shorter and its depth shallower at higher energies, an effect consistent with the X-ray emission being obscured during eclipse by a Compton-thick core of a dense wind. The spectrum is strongly disk-dominated, which allows us to constrain the spin of the black hole via X-ray continuum fitting. Three other wind-fed black-hole systems are known; the masses and spins of their black holes are high: $M\sim 10-15\msun$ and $a_*>0.8$. If the mass of IC 10 X-1's black hole is comparable, then its spin is likewise high.

\end{abstract}



\bibliographystyle{apj_fixed}


\section{Introduction}

\ic, a luminous and variable X-ray binary in the dwarf irregular galaxy IC 10, was discovered by \citet{Brandt_1997} using {\it ROSAT}.  IC 10 is notable for being the closest starburst galaxy, at a distance of $750\pm50$~kpc \citep{Demers_2004,Vacca_2007,Sanna_2008,Kniazev_2008,Kim_2009}, and for its marked overabundance of massive stars.  In particular, there is a large population of Wolf-Rayet (W-R) stars, despite IC 10 being extremely metal poor \citep[e.g.,][and references therein]{Sakai_1999,Wang_2005}. \ic\ contains one such massive W-R star that closely orbits the compact X-ray source [MAC92] 17A \citep{Clark_2004}, which we also refer to as \ic. Like a number of other X-ray binary systems (e.g., Cyg X--1, \citealt{Gallo_2005}; LMC X--1, \citealt{Pakull_1986}; SS~433,\citealt{Fabrika_2004}, XTE J1550--564, \citealt{Steiner_j1550jets,Wang_2003}; Cyg X--3, \citealt{Sanchez_2008}; GRS~1915+105 and GRO J1655--40, \citealt{Heinz_2002}), \ic\ is embedded in a low-density bubble, $\sim150$~pc across, and it is radio bright \citep{Yang_Skillman_1993,Bauer_Brandt_2004,Wang_2005}. The X-ray source exhibits a low-frequency 7~mHz quasi-periodic oscillation \citep{Pasham_2013}.  Based on the brightness of the source, in their discovery paper \citet{Brandt_1997} suggested \ic\ as a likely black hole W-R binary.

One intriguing outcome of the present census of stellar-black hole spin measurements is a possible dichotomy between the wind-fed (``X-ray persistent'') systems as compared to the Roche-lobe overflow (``X-ray transient'') systems (see \citealt{MNS14,Steiner_2014}): The transients have spins that range widely, from near-zero to near-maximal; recent evidence suggests that much or all of the spin in these systems may have been supplied through long-acting accretion torques spinning up an initially non-rotating black hole \citep{Fragos_2015}. In contrast, the three known wind-fed systems -- Cyg X--1, LMC X--1, and M33 X--7 -- harbor high-spin black holes ($\spin > 0.8$). The high spins of the wind-fed systems are especially noteworthy given the young ages of these systems, which precludes appreciable spin-up through accretion torques, implying that the spins of these black holes were imparted during the process of their formation. Another distinction between the two classes of X-ray binaries is that the black holes in the wind-fed systems are significantly more massive.  Among the three established wind-fed systems, M33 X--7, which has a massive ($M_2 \approx 70\msun$) O-star companion \citep{Orosz_M33X7}, is similar to \ic\ in that it is located in a low-metallicity Local Group galaxy at a distance of $\sim 800$~kpc and contains a quite massive $\sim15\msun$ black hole primary.

Firm dynamical estimates of the masses of two-dozen black holes (BHs) in X-ray binaries have been obtained, almost exclusively by measuring the Doppler shifts of photospheric {\it absorption} lines. Up until eight years ago, the distribution of masses was relatively narrow, $6\Msun$--$15\Msun$, a result that was upended by startling evidence, based on He II {\it emission}-line velocities, that \ic\ is comprised of a $\gtrsim30\Msun$ BH in a tight 35-hr orbit with a comparably massive $\sim 35\Msun$ W-R secondary \citep{Prestwich_2007,Silverman_2008}. Modeling the evolutionary history of this extraordinary system proved to be quite a challenge (e.g., \citealt{Bogomazov_2014}). Very recently, however, the mass estimate for IC 10 X-1 has been shown to be invalid: \citet{Laycock_2015} demonstrated that the He II line does not originate from the star, but rather from a shadowed region in the wind of the W-R companion. They conclude that the mass of the primary is currently unknown and in \citet{Laycock_2015b} that it may even be consistent with the mass of a neutron star.

 
The spins of stellar-mass BHs are presently being measured using two X-ray spectroscopic techniques: continuum-fitting and reflection modeling.  Spin\footnote{With $J$ the BH's angular momentum, $\spin \equiv cJ/GM^2$; $|\spin| < 1$.} is a quantity of great interest because according to the ``no-hair theorem'' spin and mass together uniquely and fully characterize a BH in general relativity\footnote{Electrical charge, the third defining quantity, is neutralized in astrophysical settings.}. Both methods rely upon a single foundational assumption, namely that the inner-disk terminates at the innermost stable circular orbit (ISCO). Observations establish the presence of a constant inner radius in BH systems in certain states (e.g., \citealt{Steiner_2010}).  Meanwhile, theoretical studies have identified this constant radius with the ISCO (e.g., \citealt{Zhu_2012,Kulkarni_2011,Noble_2010}, but see \citealt{Noble_2009}).

In the X-ray continuum-fitting method, one models the thermal emission from the multi-temperature accretion disk to constrain the size of the ISCO radius; the method requires accurate measurements of the BH's mass $M$, the system's inclination $i$ and distance $D$. For the reflection method, one models the relativistic distortion of fluorescent features from an accretion disk that is illuminated by a coronal source, with a focus on the $\sim 6.7$~keV Fe K line. The extended red wing is a measure of the strength of the gravitational potential and allows one to estimate the disk's inner radius. To date, application of these methods has yielded estimates of spin for a total of $\sim 20$ stellar BHs (\citealt{Reynolds_2014,MNS14,Steiner_2014}, and references therein); additionally, the spins of a comparable number of supermassive BHs have been measured via reflection modeling (see e.g., \citealt{Walton_2012,Risaliti_2013,Brenneman_2013}).


Given that the mass of the compact X-ray source is now unknown, we examine afresh the case of \ic.  We begin by considering the possibility that the X-ray source is a neutron star (NS) and show that this model is improbable. Having concluded that \ic\ contains a BH primary, we use the continuum-fitting method and an unrivaled spectrum obtained in simultaneous observations made using \chandra\ and \nustar\, to place constraints on the spin of the BH.

\section{Data}\label{section:data}

We carried out a joint observation using \chandra\ and \nustar\ for $\sim150$~ks -- just over one full orbit -- starting on UT 2014 November 6\footnote{\nustar's observing window was slightly longer at 166~ks.}.  \chandra\ was operated using a single chip, ACIS I-3, using a 100-row sub-array in order to minimize photon pileup, which causes distortion of the spectrum.  This operating mode reduced the frame-time nearly tenfold, to just 0.4~s.   Because the maximum count rate in ACIS was merely ~0.2 s$^{-1}$ (corresponding to an isotropic luminosity of $\sim 1.2 \times 10^{38}$~erg~s$^{-1}$), the resulting degree of photon pileup is minimal, $\sim 1\%$.  The quite minor impact of the remaining pileup is nevertheless accounted for in all spectral fits, using the {\sc pileup} model of \citet{Davis_pileup}.  Because the pileup was so modest, we could not fit for the model's grade migration term, and merely kept it fixed at a fiducial value of $0.7$. This and other pileup settings had no impact on our fit results, but were incorporated for completeness in the analysis.

\chandra\ data have been reduced using {\sc ciao} v4.7.  Because the data were obtained near the chip edge, the response files are calibrated using an exposure map and aperture correction, which has modest ($\sim 1\%$) impact on the effective area.  Our final spectrum employs a circular aperture with a 5~arcsec radius\footnote{We explored using a smaller, 3~arcsec aperture and results were indistinguishable.}  centered on \ic, which was near the detector aim Obpoint.  The background is obtained using a blank region from the same observation.   Data have been binned to adequately oversample the detector resolution (by a factor $\sim3$), and to a minimum of one count per bin\footnote{Necessary when employing {\sc xspec}'s c-statistic.}.  \chandra\ spectra are fitted over 0.3--9~keV (\ic\ produces insignificant signal above this range for \chandra), and \nustar\ from 3--30~keV\footnote{The upper-bound on the \nustar\ band corresponds to approximately the point at which the signal falls to $<10$\% of the background level.}.  All data were standardized to the \citet{Toor_Seward} spectral standard model of the Crab using the model {\sc crabcor} following
\citet{Steiner_2010}.  This corresponds to a shift in spectral slope of $\Delta\Gamma = 0.02$ for \chandra\ and a renormalization of $f=1.11$\footnote{This was derived comparing \chandra\ calibration data in \citet{Ishida_2011} with {\it RXTE} and {\it Suzaku}, and using the Crab calibration for those detectors from \citet{Steiner_2010}.}.  With \nustar, $\Delta\Gamma=0.00$ \citep{Madsen_2015}, and a floating cross-normalization term was fitted.

We reduced the \textit{NuSTAR} data using the standard pipeline,
\textsc{nupipeline}, part of the \textit{NuSTAR} Data Analysis
Software (\textsc{nustardas}, v1.4.1), with the latest instrumental
calibration files (caldb v20150316). The unfiltered event files
were cleaned with the standard depth correction, which
significantly reduces the internal high-energy background, and
passages through the South Atlantic Anomaly were removed. Source
spectra and instrumental responses were produced for each of the
two focal plane modules (FPMA/B) using \textsc{nuproducts},
extracted from a circular region of radius 50$''$ centered on \ic.
The background was estimated from a much larger region on the
same detector as the source. In order to maximize the good
exposure, in addition to the standard ``science'' (mode 1) data,
we also extracted the available ``spacecraft science'' (mode 6)
data. These are events collected while the source is still
visible to the X-ray optics, but the star tracker located on the
optics bench no longer gives a valid solution, so the aspect solution
is constructed from the star trackers on the spacecraft bus
instead. This typically results in some reduction in image
reconstruction capabilities, but not in the spectral response
of the instruments (see Walton et al. and Fuerst et al., in
preparation, for more details). In this case, the source point-spread function (PSF)
degradation was very mild, and so the reduction of the mode 6
data simply followed the standard procedure outlined above,
with a slightly larger source region of radius 55$''$ adopted to
account for this slight degradation. These data provided an
additional $\sim$30\% exposure for this observation. Finally,
owing to the low signal-to-noise, we combined the data from FPMA
and FPMB using \textsc{addascaspec}. The resulting
\textit{NuSTAR} spectrum provides a detection up to $\sim$30 keV.

\ic\ is in a fairly isolated field.  The closest field source is $1.0\arcmin$ from \ic, and is only 1\% as bright.  Meanwhile, the brightest source in the FOV is $<7$\% as bright and is $1.6\arcmin$ from \ic.  Neither source was problematic for the \nustar\ observations.  We note, however, that the \nustar\ data appear to imply surprisingly lower flux than \chandra.  The data are fainter than would be expected by $\sim20\%$.  In effort to understand this, we consider sources of calibration uncertainty in the \nustar\ data.  By comparing the \nustar\ point spread enclosed energy function of \ic\ to that of the model used for correcting the PSF in the \nustar\ pipeline, we estimate that the actual count rate might be off by as much as 10--15\%. In addition, \ic\ is located close to a detector edge, which at times sweeps through the PSF. This is accounted for in the pipeline, but the correction is not accurate, particularly at low count rates.  While difficult to precisely assess, a conservative 5-10\% uncertainty is likely from this. Finally, we note that the \nustar\ flux can be off globally by $\pm5\%$ \citep{Madsen_2015}.  Therefore, in total, the net error on \nustar\ fluxes may be up to $\sim25\%$.

Spectral analysis was conducted using {\sc xspec} v.12.9.0c \citep{XSPEC}, and model optimization employed {\sc xspec}'s c-statistic (``$C^2$'') \citep{Cash_1979}, as appropriate for Poisson-distributed data\footnote{We will describe the  goodness of fit in terms of $C^2$, which is analogous to the familiar $\chi^2$ statistic.}.   

\begin{figure}[h!]
\begin{center}
\includegraphics[width=0.7\columnwidth]{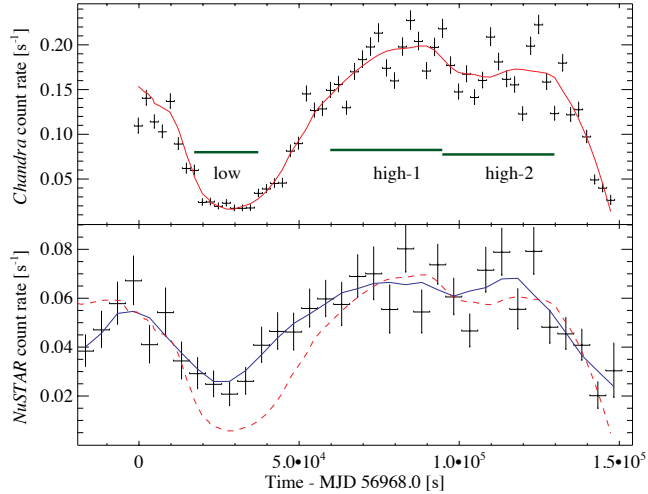}
\caption{{\it Chandra} and {\it NuSTAR} (FPMA+FPMB) background-subtracted light curves of \ic.  As a guide to the eye, a running third-order polynomial fit is shown in red over the \chandra\ data.  The out-of-eclipse (``high-1'' and ``high-2'') and in-eclipse (``low'') intervals from the text are indicated.  The impressive duration of the eclipse phase, $\sim30$\% of an orbital cycle, is a consequence of the dense and powerful stellar winds emanating from the W-R companion.  In the lower-panel, the \chandra\ curve is overlaid for reference, having been simply rescaled (red dashed).   A running fit to the \nustar\ data is in blue.  Note that the eclipse appears both shallower and also somewhat narrower in \nustar\ compared to \chandra.  This is a direct result of \nustar's sensitivity to higher energies, which are less affected by absorption in the wind.
\label{fig:lc}%
}
\end{center}
\end{figure}

The \chandra\ and \nustar\ light curves from our observation are shown in Figure~\ref{fig:lc}.  The strong dip marks the passage of the X-ray primary source behind a thick, wind-obscured core of the W-R star, marked by energy-dependent scattering (e.g., \citealt{Orosz_M33X7,Barnard_2014}).  We select in- and out-of eclipse regions (``low'' and ``high'', respectively).  The high-flux region is split into two segments (``high-1'' and ``high-2''), to check for possible phase variation as the source passes through the strong wind of the companion.  

\section{Wind and Absorption}\label{section:wind}

The broadband X-ray flux in eclipse is just $\sim$12\% of the flux out of eclipse.  Given the presence of a powerful W-R wind in which the source is embedded, the origin of this emission is likely to be electron scattering of X-rays from the photoionized wind of the W-R star and we adopt this scenario in using a corresponding model for which this signal is attributed to the inner-region's X-ray emission scattering off of an extended ``halo'' of electrons in the wind.  A schematic of this system is shown in Figure~\ref{fig:cartoon}.  As pointed out, e.g., by \citet{Barnard_2014}, this is no true ``eclipse'' in the sense of a solid body obscuring another, but rather must be attributed to a Compton-thick scattering agent, which is sensibly depicted by \citet{Laycock_2015} as a thick shell of wind surrounding the star.  This picture is similarly compatible with the parallel system NGC 300 X--1, a 33~hr orbit WR-BH binary \citep{Binder_2015}.  We therefore additionally allow for highly absorbed emission from the primary source passing through such a Compton thick wind (the scattered contribution is present in both the low and high phases); for the first time,  the high-energy coverage provided by \nustar\ allows for the detection of the transmitted photons at high energies which are insensitive to the veiling of the large gas column (as well as the Compton component).   Ideally, this model would allow for the absorption to vary with phase over the full duration of the eclipse.  However, given the faintness of the source to both instruments, we make a simplification and require a single, {\em characteristic} column of absorber to capture this effect in the ``low'' observation.  We note, however, that given sufficient signal in conjunction with a more complete model of the W-R companion and an expected wind profile, one could place a firm constraint on the inclination using the phase evolution of the column.  Our data is not of this quality, and such an investigation is beyond the scope of this paper.  

A strong and broad absorption signature is detected with \chandra\ near $\sim 2~{\rm keV}$.  The most obvious origin for such a feature is absorption in the powerful ionized winds.  We delve into the absorption features further in Section~\ref{subsec:bhspin}.  To model this appropriately, we have employed the photoionization code {\sc xstar} \citep{XSTAR} assuming the gas is hydrogen depleted (for practical reasons, we ran the code using a hydrogen abundance of 0.1 solar), and metal poor (each metal set to 0.15 solar abundance), with a solar setting for He.  We employed as an input spectrum, the average spectrum corresponding to the out-of-eclipse model, with neutral absorption removed.  We computed a table of warm absorbers corresponding to a range of columns and ionizations widely bracketing our system, spanning a range in column density of $2\times10^{20}-2\times10^{24}$~cm$^{-2}$ and of log $\xi = 0-4$.  (We assume a covering fraction of unity for the warm absorber.)  This gas is ascribed a turbulent velocity (free in the fit) using a Gaussian smearing kernel in the same manner as \citet{SWIND1}.  Here, such blurring is ad-hoc; it may, for instance, be indicative of a mixture of ionizations given the multi-phased ionization structures present in the wind (see \citealt{Laycock_2015b}, and \citealt{Vilhu_2009}).
The warm absorber's ionization is fitted for and tied across phases, but the gas column is allowed to vary between ``low'' and each of the ``high'' phases.  Because the 2~keV feature was not completely removed using this model, we also included a Gaussian absorption line that significantly improved the spectral fit ($\Delta C^2/\Delta\nu = 45.5/3$), and produced relatively minor influence on the other fit parameters.






\section{Results}\label{section:results}

We first consider models in which the compact primary of
\ic\ is an accreting neutron star with its spectrum dominated by a
thermal component of radiation from the star's photosphere. We show that
the model implies stellar radii that are much greater than those
observed for luminous accreting neutron stars, and we therefore discard
this model and consider the one viable alternative, namely, that the
compact object is a black hole. Using the continuum-fitting method,
while considering liberal ranges for the mass of the black hole and disk
inclination, we show that the black hole is spinning rapidly.

\subsection{A Neutron Star Considered}\label{subsec:ns}

Accreting NS systems in low-mass X-ray binaries fall into two categories:  ``Z'' and ``atoll'', the names of which refer to the characteristic shape traced out by each in an X-ray hardness--intensity diagram.  Between the two classes, Z-sources are the most luminous, generally emitting near, even above, the Eddington limit ($L \sim 0.5~L_{\rm Edd}$ and upwards; \citealt{Homan_2010}).   

The observed (isotropic-equivalent) source luminosity in the {\it Chandra} band is $1.5 \times
10^{38}$~erg~s$^{-1}$ (0.5--10 keV), or $\approx85$\% of the Eddington
luminosity (adopting $L_{\rm Edd} = 1.25\times 10^{38} \times
M/\msun$~erg~s$^{-1}$ and $M = 1.4~\msun$).  This luminosity is squarely
in the range observed for Z sources. 
That the spectrum of \ic\ is predominately thermal is
indicated by the much lower luminosity observed by \nustar\ above 10 keV,
which is $\lesssim5$\% of the luminosity in the full \chandra\ band.
Meanwhile, the difference between the observed luminosity and the
intrinsic source luminosity due to scattering in the stellar wind
(Section~\ref{section:wind}) is $\lesssim10$\% and unimportant for the
discussion at hand. Additional evidence that the spectrum is thermal is
provided by results presented in this section; namely, in fitting the
spectrum with thermal models, we find that the power-law component is
exceptionally faint and can be accounted for by upscattering of at most
only a few percent of the seed photons.

Before presenting our analysis of the spectrum of \ic, we discuss
pertinent results for accreting neutron stars with a focus on estimates
of their radii. The key neutron star binary for our discussion is XTE
J1701--462.  This remarkable source traced out the full-repertoire of Z
source states and then transitioned through a
lower-luminosity atoll phase \citep{Homan_2007}. In studying the
evolution of the source, \citet{Lin_2009} (hereafter L09) employed a
model comprised of three spectral components: (i) a single-temperature blackbody
from the stellar surface, which describes emission from the boundary layer (sometimes called a
``spreading'' layer); (ii) a cooler multi-temperature blackbody from an
accretion disk; and (iii) a power-law component due to Compton
up-scattering.

We note that the model of the boundary layer region is relatively
uncertain. It is frequently assumed to be 
 optically thick and of modest height, of order $\lesssim 1$~km, 
 (e.g., \citealt{Inogamov_1999,Rev_Gil_2006}).  
  Alternatively, the boundary
layer has been modeled as a hot, low-density gas that at high
luminosities can extend both radially and out of the disk plane by more
than 1 stellar radius, in which case it would behave like a scattering
atmosphere (akin to a corona; \citealt{Popham_2001}).

Given the theoretical uncertainties in the size of the boundary layer,
we turn to empirical evidence that shows it is compact.  For our
touchstone source XTE J1701--462, during its evolution as a Z source 
its power-law component was generally negligible and its blackbody radius was in
the range $\approx2-4$~km over the luminosity range $\sim0.5-2~L_{\rm  Edd}$.  
In evolving through the fainter atoll phase in soft states, the blackbody radius remained
constant at 1.7~km. In every observation throughout the outburst, the radius was 
strictly $<6$~km; such estimates are firm because the
distance estimate for XTE J1701--462 is based on super-Eddington type I
bursts \citep{Lin_2009bursts}.

Meanwhile, small radii are widely reported in studies of other luminous
accreting NSs. For the Z source GX 17+2, \citet{Lin_2012}
found that across all states the blackbody radius was $<6$~km. In a
study of six canonical Z sources, \citet{Church_2012} found blackbody
radii that are consistently $\lesssim10$~km (apart from the unstable ``flaring branch'' 
for which the maximum radius observed was 16~km). While Z sources are most
appropriate for comparison with \ic, we note that radius estimates
reported for the less luminous atoll sources are typically $\sim2$~km
(\citealt{Lin_2007,Lin_2010,Barret_Olive_2002}, and references therein)
and are strictly $<8$~km in the works cited.
{\it The small blackbody radii inferred
  for XTE J1701--462 and other NSs constitutes the strongest empirical 
  evidence for a compact NS boundary layer.}
  
We now present the results of our analysis of the data for \ic\ with a
focus on the blackbody radius of its hypothetical neutron star. Our
analysis follows closely the lead of L09. Our spectral model allows for
the modest effects due to the presence of the phase-dependent warm
absorber described in Section~\ref{section:wind} and to scattered light,
using the in-eclipse spectrum to calibrate the magnitude of its
contribution. We consider two basic models: Model 1 is a
single-temperature blackbody; Model 2 is this same blackbody component
plus an accretion disk component (modeled via {\sc ezdiskbb};
\citealt{EZDISKBB}).

For both models, we: (1) model Compton scattering of thermal photons in
a corona using {\sc simpl} \citep{Steiner_simpl}. The photon spectral
index is poorly determined, and so we fix $\Gamma=2$, a choice that is in this case
inconsequential for the determination of the blackbody radius. (2) We 
include neutral absorption using {\sc tbabs} (with {\sc wilm} abundances; \citealt{tbabs},
and {\sc vern} cross-sections; \citealt{Verner_1996}). (3) Initially, we
model the $\sim2$~keV absorption feature (Section~\ref{section:wind})
with a Gaussian. This component turns out to be significant only for
Model 1, and so we omit it in our final analysis for Model 2.

Results for the two models are summarized in Table~\ref{tab:ns}. The
data are well-fitted by both Models 1 and 2 with $C^2/\nu = 1.01$ and
0.96, respectively.  As mentioned above, the Compton component is faint,
as measured by the parameter $f_{\rm sc}$ in Table~\ref{tab:ns}; only a
few percent of the thermal photons are scattered into the power law.
The key results at the bottom of the table are the $3\sigma$
lower bounds on the blackbody radius, which were obtained using XSPEC's
error search command. {\it We establish a lower limit on the size of the
  hypothetical neutron star in \ic, including its boundary layer, of $R
  > 32.5$~km at 99.7\% confidence, which is much greater than the
  directly-comparable value of $R \lesssim 10$~km for luminous neutron
  stars discussed above.} We therefore reject a neutron star model for
\ic.

Not only does the empirical comparison of its radius against hundreds of spectra of known NS systems
rule against identifying \ic\ as a NS, but so too does interpretation of the 
radius in the context of NS models.  
We first note that emission from a neutron star's surface is physically subject to a color correction $f_c$ (a scale factor relating color temperature to effective temperature that accounts for a hot scattering atmosphere), where $f_c \gtrsim 1.3$ \citep{Suleimanov_2011}, and the emission is likewise subject to corrections that account for relativistic distortion of the emitting area (e.g., \citealt{Ozel_2013}).  Both effects serve to adjust the size of the emitter {\em upwards}, in the sense that the true size would be strictly larger than implied when such effects are ignored.  Similarly, any obscuration by the disk would likewise serve to increase the true size of the emitter when compared to the blackbody fit result.  Therefore, the size returned by the simple and non-relativistic blackbody model, which neglects these corrections, is already guaranteed to underestimate the true size.  
This is important given that the lower limit on the boundary layer size of 32.5~km is already large compared to the maximum NS size ($\lesssim 16~$km, e.g., \citealt{Gandolfi_2012}).    

Finally, we consider the physical He-atmosphere model NSX of
\citet{NSX}, which includes the effects of spectral hardening and
relativistic distortion. Replacing directly the blackbody component with
this model component and repeating our analysis, we obtain the results
for Models 1a and 2a, which are given in the rightmost columns of
Table~\ref{tab:ns}. For the NSX model, our lower limit on the radius is
$R > 250 $~km at 99.7\% confidence, which is far greater than predicted
by any theoretical model of a neutron star.

In summary, using a simplistic blackbody component to describe the boundary layer emission, we have established a hard lower limit on the emitting surface area which is still a factor $2-3$ larger than has been observed for any accreting neutron star in, respectively, flaring or stable states.  We are therefore forced to conclude that either \ic\ has unique properties among the NSs which cause it to appear so large, or else it is a BH.  While we cannot definitively rule out the possibility that \ic\ is an exotic NS, parsimony via Occam's razor supports its identification as a BH.

Therefore, we rule against the possibility of a neutron star primary, and with the black-hole nature of \ic\ prevailing, now change focus to black hole spectral models.

\begin{deluxetable}{lcccc}
  \tabletypesize{\scriptsize}
  \tablecolumns{5}
  \tablewidth{0pc}
  \tablecaption{Spectral Fit Assuming a Compact X-ray Star}
  \tablehead{ Model  &  {\sc blackbody}  &  {\sc blackbody} + {\sc ezdiskbb}  &   {\sc nsx}  &  {\sc nsx} + {\sc ezdiskbb}   \\
                     &       (1)         &       (2)                          &      (1a)    &        (2a)
} \startdata
$N_{\rm H} (10^{22}~{\rm cm}^{-2})$           & $0.32\pm0.07$          &  $0.58 \pm 0.12$       & $0.42 \pm 0.07$   & $0.20\pm0.09$  \\
$\Gamma$                                      & 2                      & 2                      & 2                 & 2    \\
$f_{sc}$                                      & $<0.01$                & $0.040 \pm 0.015$      & $<0.02$           & $<0.01$         \\
$kT_{\rm ezdiskbb}$ [keV]                     & \nodata                & $0.30\pm0.04$          & \nodata           & $0.20 \pm 0.01$      \\
$N_{\rm ezdiskbb}$                            & \nodata                & $3.5^{+4.0}_{-2.0}$    & \nodata           &    $800_{-250}^{+350}$  \\
$\tau_{\rm sc. em.}$                          & $0.12\pm0.01$          & $0.09\pm0.01$          & $0.12 \pm 0.01$   &   $0.13\pm0.01$ \\
$N_{\rm NuSTAR}/N_{\rm Chandra}$              & $0.81\pm0.05$          & $0.80\pm0.05$          & $0.81\pm0.05$     &  $0.80\pm0.05$  \\
$kT_{\rm star}$ [keV]                         & $0.50\pm0.08$          & $0.88 \pm 0.03$        & \nodata           & \nodata   \\     
log ($T$/[K])$_{\rm nsa}$                     & \nodata                & \nodata                & $6.60\pm0.06$     &  $6.7_{-0.005}$  \\     
$R_{\rm star}$  [km]                          & $1600^{+10000}_{-1300}$&  $38.8 \pm 3.4$        &  $1500_{-1000}^{+1500}$ &  $260\pm6$   \\
$R_{\rm star}$ Lower Bound ($3\sigma$) [km]   & $>155$                 &   $>32.5$              & $>330$            &  $>250$  \\
\hline
$C^2$ / d.o.f.                                & 792.7 / 781            &    752.6/782           & 790.3/781         &   827.0/ 782 \\
\enddata
\tablecomments{For clarity, this table omits the extraneous warm-absorber parameters.  Those values are in line with fits presented below in Section~\ref{subsec:bhspin}, and are inconsequential in determining $R_{\rm star}$.  All uncertainties are 1$\sigma$ equivalent confidence intervals, except as noted for the hard limit on $R_{\rm star}$.}
\label{tab:ns}
\end{deluxetable}




\subsection{Black Hole Spin Via Continuum Fitting}\label{subsec:bhspin}

Spin is determined by fitting a disk-dominated spectrum using the
relativistic thin-disk model of Novikov \& Thorne
 (\citeyear{NT73}). In practice, one uses the publicly-available  
 codes {\sc kerrbb} \citep{KERRBB} and {\sc bhspec} \citep{BHSPEC}.  The model determines the inner edge of the BH's
 accretion disk $R_{\rm ISCO}$, which is trivially related to the spin
 parameter $a_*$.  
 Following the procedure of \citet{Steiner_2012_H1743}, we employ {\sc kerrbb} and {\sc bhspec} to generate a table of color correction values across a grid of $M$, $i$, $\spin$, and luminosity, customized to the spectral responses of \chandra\ and \nustar.  This table is then employed by our standard model package, {\sc kerrbb2} (\citealt{McClintock_2006}).  

 A Compton power-law component is generated using {\sc simpl}, which mimics the behavior of a corona, scattering a fraction of disk photons ($f_{\rm sc}$) into a power law with photon index $\Gamma$ \citep{Steiner_simpl}.  This contribution turns out to be quite weak.  Line-of-sight absorption through the interstellar medium is calculated using {\sc tbabs}.  Absorption which is most prominent below 2~keV has been modeled as a warm absorber associated with the strong stellar wind, and an additional Gaussian was included as well, as described in Section~\ref{section:data}.   The warm absorber is allowed to vary in column density with orbital phase.
 Inclusion of these absorption features improves the fit significantly, by $\Delta\chi^2/\Delta\nu = 53/9$, and causes a $\sim 15\%$ increase in the fitted inner radius (so that omitting absorption from the model results in a higher spin).

\begin{figure}[h!]
\begin{center}
\includegraphics[width=0.7\columnwidth]{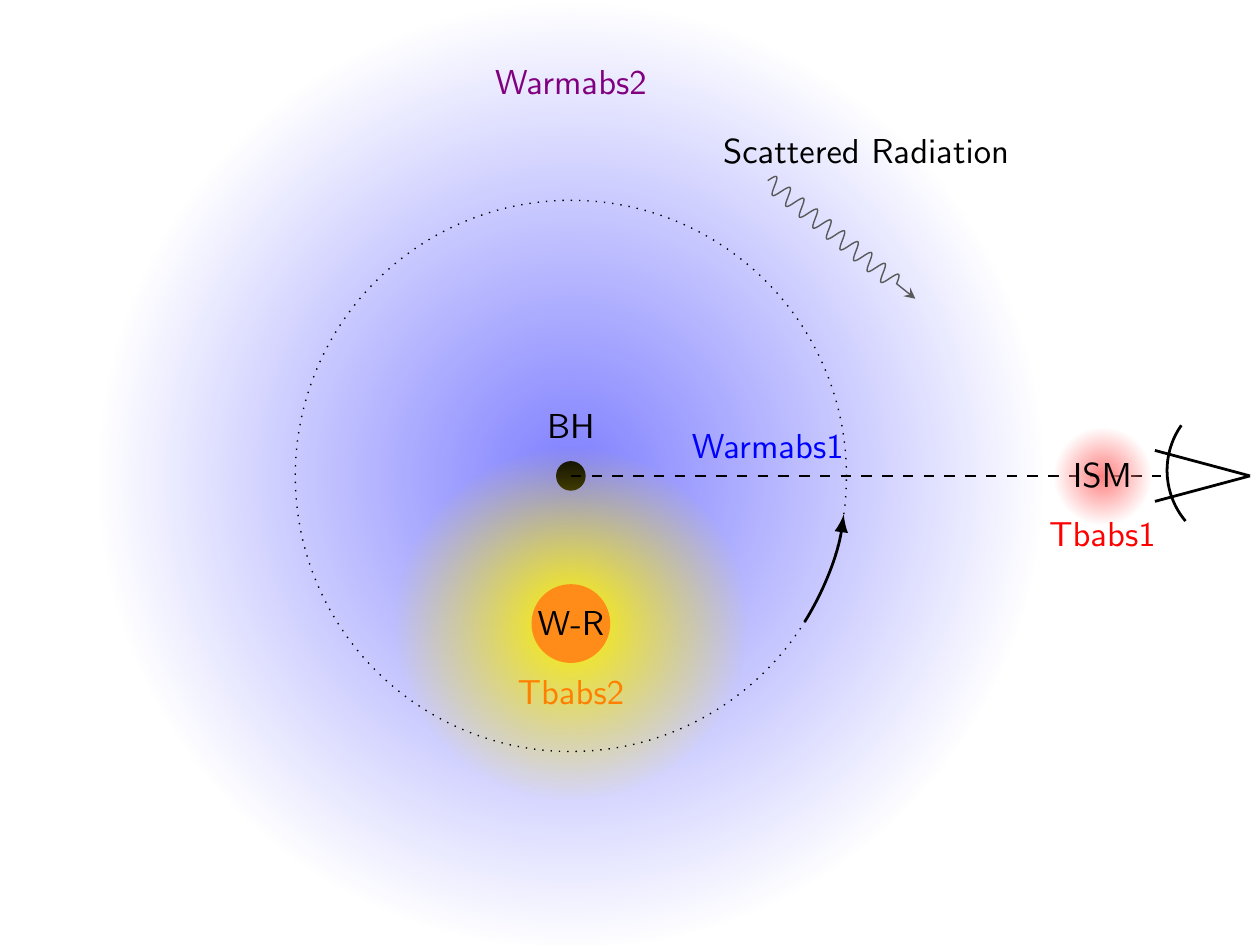}
\caption{A schematic of the BH -- W-R system illustrating the various wind and warm absorber components.  \label{fig:cartoon}%
}
\end{center}
\end{figure}

\begin{figure}[h!]
\begin{center}
\includegraphics[width=0.7\columnwidth]{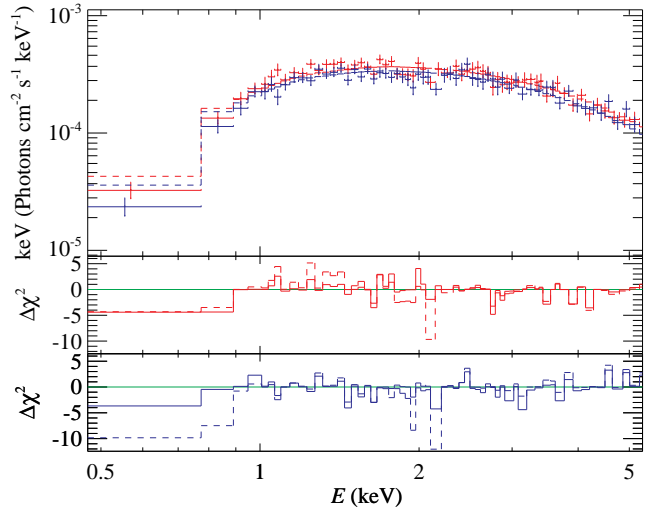}
\caption{The effect of absorption on the spectral fits.  Here, we zoom in on the low-energy region of the ``high'' \chandra\ data, at which absorption effects are evident, and we present model fits without any absorption included (dashed lines).  The bottom panels show the resultant $\chi^2$ residuals with the sign of data-minus-model and the evident fit improvement --particularly around 2~keV-- when absorption is incorporated into the model (solid lines) for each of the ``high'' spectra in turn.  The inclusion of absorption increases the intrinsic flux of the underlying disk component in the fit, which has the effect of lowering $R_{\rm in}$, and accordingly increasing the inferred spin. \label{fig:abschi}%
}
\end{center}
\end{figure}

 Our full, composite model is fitted to the six spectra (each of \chandra\ and \nustar\ for the three phase intervals) at once, and is structured as: {\sc crabcor $\times$ pileup(gabs $\times$ tbabs1 $\times$ [  warmabsorber1 $\times$ tbabs2 (simpl $\otimes$ kerrbb2) + warmabsorber2 $\times$ const (simpl $\otimes$ kerrbb2)])}.  Here, the constant term determines the contribution of X-ray emission scattered into our line of sight by the halo of electrons in the extended stellar wind.  {\sc tbabs1}\footnote{The neutral absorption ({\sc tbabs1}) was assumed to be Galactic in origin, and is in line with the Galactic column of $N_{\rm H} \sim 4-5\times10^{21}{\rm cm^{-2}}$\citep{Dickey_Lockman}.  However, as noted in \citet{Barnard_2014}, a modest improvement in the fit is observed when using the lower metal abundances of IC~10.  Given that the full neutral column is expected from the Galactic contribution alone, we have opted against using this lower abundance fit here.  The difference in goodness is $\Delta C^2 = 15$, and the spin and other parameters of interest are unaffected within errors. The sole difference is in the hydrogen-equivalent column density, which changes by a factor $\sim 4$.} represents neutral absorption along the line of sight whereas {\sc tbabs2} gives absorption in the thick shell of wind during eclipse;  accordingly, it is a fit term in the eclipsing ``low'' phase but the column is otherwise fixed to zero.  {\sc warmabsorber1} describes the phase-dependent absorption by the wind of the source, while the column for {\sc warmabsorber2} is linked among all phases and describes the attenuation by absorption in the wind for the diffuse, scattered light.  All parameters in {\sc simpl $\otimes$ kerrbb2} are linked between their two instances.  (All warm absorber terms are linked to a single ionization parameter and turbulent velocity width, which turn out to be poorly constrained.)  The  illustration in Figure~\ref{fig:cartoon} shows the correspondence of these components to the structure of the system.  The warm-absorber's influence on the fit is presented in Figure~\ref{fig:abschi} for the two ``high'' \chandra\ spectra.
 
Based on the similarity between \ic\ and the eclipsing high-mass black-hole binary M33 X--7, we adopt round fiducial values for the black-hole mass and inclination of $15~\msun$ and  $75\degr$, respectively.  Later, we will examine the mass-and-inclination dependence of our results.  The fits, meanwhile, are only weakly sensitive to the value of the disk-viscosity term ($\alpha$), and here we pick a reference value of $\alpha = 0.05$.  We allow for differences with phase, including in $\mdot$ between intervals ``high-1'' and ``high-2'' (in the ``low'' interval, we match $\mdot$ to that obtained from ``high-1'')\footnote{Opting instead to use the $\dot{M}$ value from ``high-2'' is inconsequential.}.  Aside from the warm absorber column and $\mdot$ terms, all other parameters are tied among the observations.  

\begin{figure}[h!]
\begin{center}
\includegraphics[width=0.7\columnwidth]{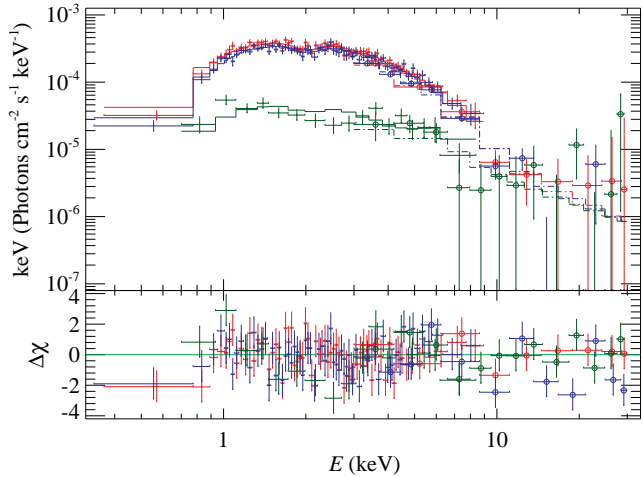}
\caption{The best-fitting model and accompanying residuals to our joint {\it Chandra} and {\it NuSTAR} data set.  The data have been rebinned for plotting purposes only.  The three time intervals marked in Fig.~\ref{fig:lc} are fitted jointly.  Fit parameters are given in Table~\ref{tab:fit}.  ``High-1'' and ``high-2'' are shown in red and blue, respectively, and green shows the ``low'' data for both \chandra\ and \nustar.  \nustar\ data are marked with open circles, and the \nustar\ spectral models are given with dot-dashed lines, whereas \chandra's model fits are solid lines.
\label{fig:fit}%
}
\end{center}
\end{figure}

We optimize the fit initially by directly fitting in {\sc xspec}, and then for a more robust analysis a full exploration of the model is performed via Markov-Chain Monte Carlo (MCMC) using the {\sc emcee} algorithm \citep{emcee} following the setup described in \citet{Steiner_lmcx1}.  
Here, we apply usual noninformative priors to nearly all terms, either uniform in linear space for shape parameters (such as $\spin$ and $\Gamma$), or uniform on the logarithm for terms with no preferred scale (such as $N_{\rm H}$ or the normalizations).  The single informative prior is on the {\sc crabcorr} normalization for \nustar, which sets its cross-normalization relative to \chandra.  We use a Gaussian centered on unity, with a width $\sigma = 0.1$, based upon \citet{Madsen_2015}.  
From experience gained in \citet{Garcia_gx339}, we favor employing a smaller number of walkers (we use 100, for 18 free parameters), in favor of running longer chains.  Here, we run each walker for 50,000 steps, and discard the first 10,000 steps of each as burn-in.  The auto-covariance of the parameters is significant, generally several hundred steps.  Our report is comprised of the final 4 million aggregate steps, which amounts to many thousands of independent samplings for each parameter (the hundred-fold reduction is a result of the long-lived autocovariance in the chains).  The best fitting black-hole model, which achieves $C^2/\nu = 1.00$, is summarized in Table~\ref{tab:fit} and shown in Figure~\ref{fig:fit}.  The errors reported are (minimum-width) 90\% confidence intervals.  Notably, as was observed for M33 X--7, the system appears {\em remarkably} thermal, with scant allowance for nonthermal contribution from a corona.  The spin obtained for this reference model is rather high and reasonably-well constrained, $\spin = 0.85^{+0.04}_{-0.07}$ (90\% confidence), where the error reflects {\em statistical} uncertainty only.

\begin{figure}[h!]
\begin{center}
\includegraphics[width=0.7\columnwidth]{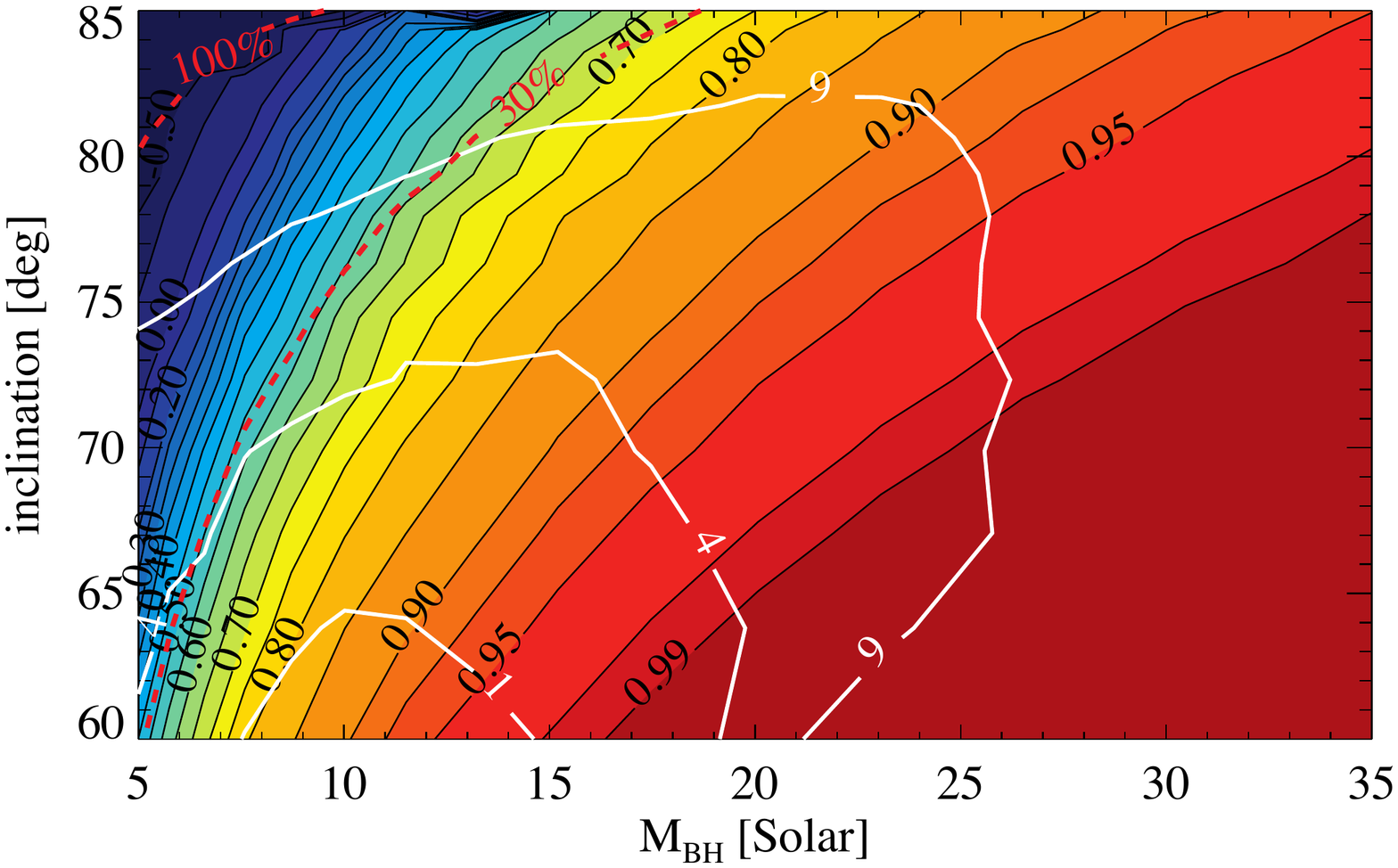}
\caption{Spin as a function of the BH's mass and inclination.  $\Delta C^2$ contours are overlaid in white, marking out the weakly preferred values of mass, spin, and inclination.  The statistical uncertainty in spin, while not shown here, scales nonlinearly in spin, but is rather constant in units of $\Delta R/R_{\rm ISCO}$ (the $1\sigma$ error being $\approx 5.5\%$).  The red dashed lines show contours of constant luminosity, in percentage of the Eddington limit.  The lower dashed line at 30\% Eddington separates the region at which the model is within the thin-disk limit (below and to the right of the line) from the region for which the disk would be appreciably thick.  
\label{fig:contour}%
}
\end{center}
\end{figure}

\begin{figure}[h!]
\begin{center}
\includegraphics[width=0.7\columnwidth]{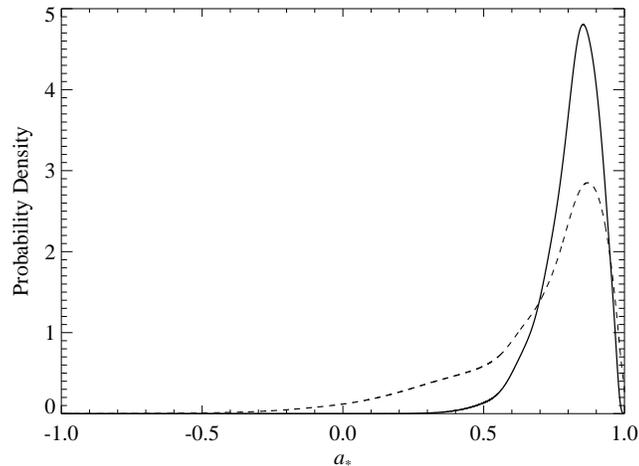}
\caption{A probability distribution estimate for spin, in which fit results have been weighted according to the goodness of fit.  The solid line shows the spin distribution associated with BH masses in the wind-fed range, $10-15~\msun$, and allowing for the full range of allowed inclinations.  However, secondary model parameters (e.g., $\alpha$ and $D$) have been merely fixed at their nominal values.  The dashed line shows a complementary constraint, in which we crudely assess a broad range of uncertainty by allowing for many unknowns, e.g., we consider the full range of explored BH masses, from $5-35~\msun$, again consider all allowed inclinations, and similarly incorporate the uncertainty in $D$ and the full span of possible settings for $\alpha$.  
\label{fig:spinpdf}%
}
\end{center}
\end{figure}

To elucidate the dependence of the spin constraint on the system's mass and inclination, we have repeated our fit over a grid in mass and inclination, shown in Figure~\ref{fig:contour} (we have separately explored varying distance and $\alpha$; Section~\ref{section:discussion}).  A firm lower bound on the system inclination $i > 63\degr$ is possible due to the strong eclipse \citep{Laycock_2015}. Likewise, the fact that the system is X-ray persistent means that disk self-shadowing cannot be substantial, which precludes extremely high inclinations comparable to the disk scale height $h/r \sim 0.1$ (i.e., $i \lesssim 83\degr$).  We note that while the spin is strongly degenerate with changes in $M$ and $i$, there is nevertheless some sensitivity to the shape of the continuum for a given inclination and spin. There is a modest preference among the data for a ``typical'' BH mass $M\sim5-20~\msun$ (e.g., \citealt{Ozel_2010}), and for the inclination to be lower $i \lesssim 70\degr$. This can be read from the white contours which are overlaid showing iso-surfaces of $\Delta C^2$ (measured relative to its global minimum). We note that the masses of the handful of wind-fed BHs are high ($\sim 10\msun-15\msun$) compared to transient BH systems (\citealt{Ozel_2010}).  Across that mass range, the spin determination for \ic\ is generally high, with $\spin \gtrsim 0.8$ for most of the interval. This spin constraint for the wind-fed BHs mass range is illustrated in Figure~\ref{fig:spinpdf}.  We have assumed a prior probability on each setting of $1/M$ (which has the effect of shifting spin toward lower values), as a naive proxy for a BH mass distribution which would favor lower masses, and apply a weight to each fit result according to its goodness ($w_i = \exp (-C^2/2)$).  The dashed line in this figure illustrates a rough estimate of the effects of considering both a broad mass range and systematic uncertainty on the spin constraint.  
We note that if, despite its dubious standing, the large mass obtained by \citet{Silverman_2008} of $M > 30~\msun$ is later borne-out, then the corresponding spin must be extreme.




\begin{deluxetable}{lccccc}
  \tabletypesize{\scriptsize}
  \tablecolumns{6}
  \tablewidth{0pc}
  \tablecaption{IC 10 X--1 Black Hole Spectral Fits}
  \tablehead{Parameter &  ``Global'' Setting & High-1 & High-2 & Low   \\
} \startdata
$N_{\rm H} <{\rm tbabs1}> [10^{22} {\rm cm}^{-2}]$ &                            $0.45\pm0.06$ &                      \nodata &                     \nodata &                     \nodata & \\
warm-abs column $<$warmabs1$>$ [$10^{22} {\rm cm}^{-2}]$ &        $<1.6$ & $<1.3$ &    $1.3-25.$ &                      \nodata & \\
warm-abs column $<$warmabs2$>$ [$10^{22} {\rm cm}^{-2}]$ &                           \nodata &                     \nodata &                     \nodata &  $<80$ &  \\
$\Gamma$ &                                                       $2.5^{+0.5}_{-0.7}$ &                      \nodata &                     \nodata &                     \nodata & \\
$f_{\rm sc}$ &                                              $<10\%$     &                \nodata &                     \nodata &                     \nodata & \\
$a_*$ &                                                          $0.85^{+0.04}_{-0.07}$ &                      \nodata &                     \nodata &                     \nodata & \\
$\dot{M} [10^{18}$ (g/s)] &                                                      \nodata &        $2.4^{+0.7}_{-0.4}$ &         $2.3^{+0.7}_{-0.4}$ &                      \nodata & \\
$N_{\rm H} ({\rm shell}) <{\rm tbabs2}>  [10^{22} {\rm cm}^{-2}]$ &                          \nodata &                     \nodata &                     \nodata &         $130^{+55}_{-40}$ &  \\
$\tau_{\rm sc. em.}$ &                                                  $0.124\pm0.013$ &                      \nodata &                     \nodata &                     \nodata & \\
warm-abs log $\xi$ &                                                $1.3^{+2.0}_{-0.3}$ &                      \nodata &                     \nodata &                     \nodata & \\
warm-abs log ($v_{\rm turb}$/c)  &                          $<0.5$                     \nodata &                     \nodata &                     \nodata & \\
Norm$_{\rm Toor \& Seward, NuSTAR}$ &                                     $0.81\pm0.05$ &                      \nodata &                     \nodata &                     \nodata & \\
$M   [\msun]$ &                                                                     15* &                      \nodata &                     \nodata &                     \nodata & \\
$i$  [degrees] &                                                                    75* &                      \nodata &                     \nodata &                     \nodata & \\
$D$  [kpc] &                                                                       750* &                      \nodata &                     \nodata &                     \nodata & \\
$\alpha$ &                                                                        0.05* &                      \nodata &                     \nodata &                     \nodata & \\
$\Delta \Gamma_{\rm Chandra}$ &                                                   0.02* &                      \nodata &                     \nodata &                     \nodata & \\
Norm$_{\rm Toor \& Seward, Chandra}$ &                                            1.11* &                      \nodata &                     \nodata &                     \nodata & \\
$\Delta \Gamma_{\rm NuSTAR}$ &                                                     0.0* &                      \nodata &                     \nodata &                     \nodata & \\
$E_{\rm gabs}$ [keV]   &                                                  $2.02\pm0.09$ &                      \nodata &                     \nodata &                     \nodata & \\
$\sigma_{\rm gabs}$  [keV] &                                     $0.20^{+0.13}_{-0.09}$ &                      \nodata &                     \nodata &                     \nodata & \\
$N_{\rm gabs}$   &                                               $0.12^{+0.08}_{-0.05}$ &                      \nodata &                     \nodata &                     \nodata & \\
pileup g0 &                                                                        1.0* &                      \nodata &                     \nodata &                     \nodata & \\
pileup alpha &                                                                     0.7* &                      \nodata &                     \nodata &                     \nodata & \\
pileup psffrac &                                                                  0.95* &                      \nodata &                     \nodata &                     \nodata & \\
\hline
$C^2/\nu$ &                                                        773.43/774 &                      \nodata &                     \nodata &                     \nodata & \\
\enddata
\tablecomments{All ranges are 90\% confidence intervals. Starred values were frozen in the fit.}
\label{tab:fit}
\end{deluxetable}

\section{Discussion}\label{section:discussion}

Our thin-disk model {\sc kerrbb2} delivers reliable estimates of spin only at luminosities $L \lesssim 30\% L_{\rm Eddington}$ and for spectra that are disk-dominated \citep{MNS14}.  The spectrum of \ic\ amply meets these two criteria for values of $M$ characteristic of the wind-fed BHs ($\sim 10-15\msun$) and for the favored range of $i$ ($\sim 65-75\degr$).  However, at sufficiently low mass and high inclination the luminosity exceeds 30\% of Eddington, as indicated in Figure~\ref{fig:contour} by the red-dashed contour labeled ``30\%''. In this regime, our model underpredicts the spin and our results become increasingly unreliable with increasing luminosity (see, e.g., \citealt{Straub_2011}).  (We note that here we use the usual definition of the  Eddington luminosity, namely $L_{\rm Edd} = 1.3 \times 10^{38} {\rm erg~s^{-1}} (M/\msun)$, which describes the luminosity at which radiation in an isotropic hydrogen sphere is at equilibrium with gravity.  However, the corresponding luminosity at which this equilibrium is reached for a He atmosphere can be a factor of 2 higher.)   

Our spin estimates rely on a standard version of {\sc bhspec} which assumes solar metallicity. While this is not ideal, it is likely a reasonable value given that the depressed metallicity of IC 10 ($\sim10-20$\% that of the Galaxy) is compensated by the hydrogen-depleted atmosphere of the W-R companion \citep{Clark_2004}. We assessed this source of uncertainty using alternate {\sc bhspec} models with metallicites that are 0.1 and 0.5 the solar value and found that the effect on our results is negligible. We similarly explored the effect of varying $\alpha$; switching between the two default values of 0.01 and 0.1 has a $\sim 3-5\%$ effect on $R_{\rm ISCO}$. The uncertainty in the distance has the biggest effect, resulting in an error in $R_{\rm ISCO}$ of $\sim 6\%$, which is comparable to the precision of the spectral fit. In sum total, the systematic uncertainty in the spin constraint is $\sim 10\%$ on $R_{\rm ISCO}$; for reference, this is equivalent to $\sigma_{\spin} = 0.045$ at $\spin = 0.85$.



\begin{figure}[h!]
\begin{center}
\includegraphics[width=0.7\columnwidth]{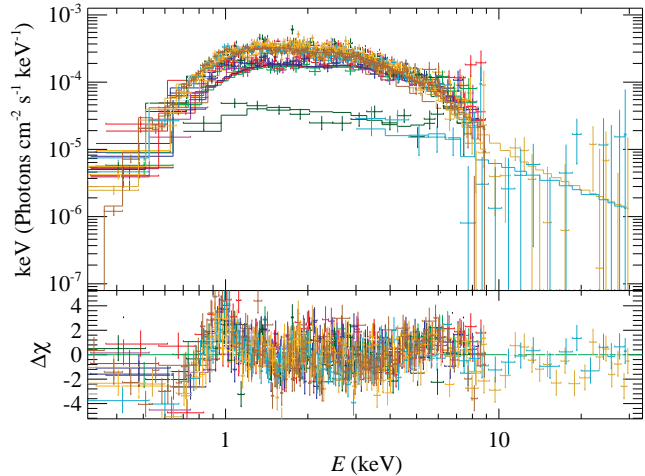}
\caption{As in Fig.~\ref{fig:fit}, but including a dozen other {\it Chandra} and {\it XMM-Newton} spectra.  These spectra span a factor $\sim 2$ in luminosity. Although the other data lack high energy coverage, we find they are quite consistent with our broadband fit for spin.
\label{fig:multifit}%
}
\end{center}
\end{figure}

In addition to our November 2014 observation, \chandra\ made three $30-50$~ks observations of \ic\ as well as several 15~ks snapshot observations. The frame times were longer for these observations than for ours and 10--20\% of the events are piled up. \xmm\ observed \ic\ on two occasions with exposure times of 45 and 135~ks. We analyzed all of these out-of-eclipse data. Figure~\ref{fig:multifit} shows fits to the \chandra\ and \xmm\ spectra, along with our data.  Only the \chandra\ data were corrected for pileup. All the system parameters were tied except for $\dot{M}$, $f_{\rm sc}$, and the column of {\sc warmabsorber1}, which were allowed to vary from observation to observation. Apart from the November 2014 data, the constraints on $f_{\rm sc}$ are weak because of the lack of high-energy coverage.

Note that the source luminosity from all epochs spans a factor $\sim 2$ (Figure~\ref{fig:multifit}), in line with the range observed in other wind-fed X-ray binary systems (e.g., Cyg X--1). Unfortunately, none of the spectra, apart from the November 2014 observation, can deliver an independent and reliable estimate of spin primarily because of the limited bandwidth, which does not allow one to isolate the thermal component from the Compton component. For example, consider the 135~ks \xmm\ spectrum which has the highest $S/N$. Fits to this spectrum allow a broad range for the normalization of the Compton component, $f_{\rm sc}=0.02-0.32$ ($1\sigma$ level of confidence).

We can nevertheless determine for the complete collection of spectra that the spin is very similar to the value we obtained in Section 4.2 by analyzing just the \chandra\ plus \nustar\ spectra.  A joint fit to all the spectra in Figure~\ref{fig:multifit} yields $\spin = 0.76 \pm 0.03$ where we have assumed that scattered light is a constant fraction (set by the ``low'' spectrum) of the disk emission.  We obtain a poorer fit in this case, with $C^2/\nu = 1.44$.  While it is possible that variation in the wind scattering or inadequate modeling of pileup may be responsible for the poorer fit and marginal decrease in spin, we note that the luminosities of several of the spectra exceed our nominal limit of 30\% of Eddington, which can depress the spin value.


\ic\ has characteristics very similar to the other eclipsing BH system in the Local Group, M33 X--7, which is
comprised of a $70\Msun$ O-giant and a $\sim15\Msun$ BH
\citep{Orosz_M33X7}.  For this 3.5-day period system, the duration of the X-ray
eclipse, including the effects of the O-star's extended wind ($\dot M_{\rm wind}
\approx 2.6{\times}10^{-6}M_{\odot}$~yr$^{-1}$), is 0.15 in phase, with the 
full width and speed of ingress strongly influenced by the absorption and scattering of
X-rays in the wind.  

As was explored for M33 X--7 \citep{Orosz_M33X7}, and as considered for \ic\ by \citet{Laycock_2015b}, we compute the Bondi-Holye-Littleton (BHL) rate of mass capture by \ic\ in order to check its compatibility with the mass supply and efficiency of the BH ($\mdot \approx 10^{18}-10^{19}$~g~s$^{-1}$ from our fits over the allowed range of $M$ and $i$ in Figure~\ref{fig:contour}). From \citet{Edgar_2004}, for a supersonic wind,
\begin{equation}
\dot{M}_{\rm BHL} = \frac{4 \pi G^2 M^2 \rho}{v^3},
\label{eqn:bondihoyle}
\end{equation}
where $v$ is the relative speed of the wind and $\rho$ is the wind's density.  From \citet{Clark_2004}, we adopt $v_{\infty} = 1750$~km~s$^{-1}$, and use the approximate scaling law:
\begin{equation}
v(r) = v_{\infty}\left(1-\frac{R}{r}\right)^{\beta},
\label{eqn:vwind}
\end{equation}
\citep{Crowther_2007}, where $R$ is the stellar radius and $\beta \approx 1$.  At the binary separation $a \sim 20 R_\Sun$ \citep{Laycock_2015b} we estimate roughly $v \approx 1600$~km~s$^{-1}$, which includes the moderate effect of orbital motion (appropriate for the range of typical BH masses).  Using the mass-loss rate from \citet{Clark_2004} of $\dot{M}_{\rm wind} = 10^{-5}~\msun~{\rm yr}^{-1}$, the predicted electron and mass densities in the helium-dominated wind are $n_e \approx 4\times10^{10} {\rm cm}^{-3}$, and $\rho \approx 2\times 10^{-13}$~g~cm$^{-3}$.  For the range of BH masses in question $\sim 5-40\msun$, $\mdot_{\rm BHL}$ comes out in the range $2\times 10^{17}-2\times 10^{19}$~g~s$^{-1}$, with larger values corresponding to solutions for higher BH masses.  Although the estimate is crude, it demonstrates that mass capture and subsequent accretion from the W-R wind is fully capable of powering the BH.

We can check on the above prediction for the wind profile by noting that the change in the line-of-sight column of the warm absorber, while weakly constrained, grows over the orbit by, very roughly $\sim 10^{22}$~cm$^{-2}$ (noting that as determined from the MCMC analysis, the increase is at a mere 2$\sigma$ significance), being largest when eclipsed.  Because the source has reached a pathlength greater by $\sim a$, the density in the wind is $n_e \approx \Delta N / a \approx 10^{10}~{\rm cm}^{-3}$.  When accounting for the abundance differences between IC 10 and the Galaxy, the absence of H in the W-R star, again the dearth of metals wash out to a correction factor of roughly unity.
This bolsters the above picture for mass-loss in the system and underscores the assumed association between warm absorber and the W-R wind.  Finally, we observe that a simple estimate for the ionization pattern far from the BH at high latitudes is that the ionization parameter $\xi \sim L/nr^2$ should be of order thousands, and that given sufficient resolution one would expect to find a dark cone with a factor $\sim 2$ lower ionization in the shadow of the star.  In practice, the constraints on the ionization parameter from the observations are weak, and so we can only say that our results are not in conflict with this value.

\section{Conclusions}\label{section:concs}
  
In summary, using a \chandra/\nustar\ observation we have demonstrated that the compact primary in \ic\ is implausible as a neutron star and therefore a black hole explanation is highly likely. Although the mass is uncertain, for any pairing of black hole mass $M$ and inclination $i$ we have determined a unique and precise estimate of the spin parameter $a_*$ using the continuum-fitting method. The strongly disk-dominated spectrum of \ic\ makes this method an especially reliable one. Meantime, this distant source ($\sim750$~kpc) is presently too faint for application of the Fe-line method.  {\it NuSTAR's} high-energy coverage picks up precisely where {\it Chandra's} effective area is falling.  Critically, the mutual capabilities of \chandra\ and \nustar\ allow one to firmly anchor the power-law component and thereby isolate and reliably model the thermal component.

Our excellent data set allows us to obtain a net precision of $\sigma_{\spin} \approx 0.05$.  However, we are hampered by a serious limitation, namely the uncertain mass of the black hole. We meet this challenge by computing the spin as a function of $M$ and $i$ over a broad range of these parameters; our constraints on $a_*$ are displayed in Figure~\ref{fig:contour}.  We find that if the mass is comparable to that in the other wind-fed systems (a value significantly above the typical mass of a transient black hole) then the spin of \ic\ is likely be high ($\spin \gtrsim 0.7$), as it is for the other wind-fed systems.

The massive W-R companion implies a young age for \ic, which in turn implies that the high spin was imparted to the black hole during its birth event. It is important now to attempt to estimate the mass of the black hole because the combination of high-mass and high-spin would strengthen the apparent dichotomy between wind-fed and transient black hole X-ray binaries (Section 1), which would point to a distinct formation channel for BH systems with massive companions (e.g., \citealt{Kochanek_2015}). Presently, while we can rule against a neutron-star primary, the jury is still out on the mass -- and hence the spin -- of the BH in \ic. If in future work the mass of the BH can be constrained, then our results provide an immediate complementary estimate of its spin.

  \acknowledgments

  This research has made use of software provided by the Chandra X-ray
Center (CXC).  This work was made possible by Chandra Grant GO4-15051X.  
JFS has been supported by the NASA Hubble Fellowship grant HST-HF-51315.01, and the NASA Einstein Fellowship grant PF5-160144.   
We thank the CXC helpdesk and Larry David for their advice on the \chandra\ 
data reductions.  JFS thanks R. Remillard and J. Homan for helpful discussions on NS spectral models.  We thank the anonymous referee for helpful criticisms which have improved this paper.  This research has made use of data obtained with the \textit{NuSTAR} mission, a project led by the California Institute of Technology (Caltech), managed by the Jet Propulsion Laboratory (JPL) and funded by NASA, and has utilized the \textit{NuSTAR} Data Analysis Software (\textsc{nustardas}) jointly developed by the ASI Science Data Center (ASDC, Italy) and Caltech (USA).

  \dataset [ADS/Sa.CXO#obs/15803] {Chandra ObsId 15803}

\end{document}